\documentclass[usegraphicx,usenatbib]{mn2e}
\pdfoutput=1

\usepackage{graphicx}
\usepackage{color, graphics}
\usepackage{amsmath,amssymb,amsfonts}
\usepackage{subfigure}  
\usepackage{acronym}
\usepackage{mathrsfs}
\usepackage{multirow}
\usepackage{url}
\usepackage{hyperref}
\usepackage{nicefrac}
\def\beq{\begin{equation}\begin{aligned}}
\def\eeq{\end{aligned}\end{equation}}

\newcommand{\DM}{\text{DM}}
\newcommand{\SM}{\text{SM}}

\newcommand{\keV}{\text{ keV}}

\newcommand{\Pcl}{P_{\rm cl}}

\title{Cores in Dwarf Galaxies from Fermi Repulsion}

\author[Lisa~Randall, Jakub~Scholtz, and James Unwin]{Lisa~Randall,$^{1}$ Jakub~Scholtz,$^{1}$ and James Unwin$^{2}$\\
{}\\
$^1$ Department of Physics, Harvard University, Cambridge, MA 02138, USA\\ 
$^2$ Department of Physics,  University of Illinois at Chicago, Chicago, IL 60607, USA}

\begin{document}

\maketitle

\begin{abstract} 

We show that Fermi repulsion can lead to cored density profiles in dwarf galaxies for sub-keV fermionic dark matter.  We treat the dark matter as a quasi-degenerate self-gravitating Fermi gas and calculate its density profile assuming hydrostatic equilibrium.  We find that suitable dwarf galaxy cores of size $\gtrsim 130$~pc can be achieved for fermion dark matter with mass in the range 70~eV~--~400~eV. While in conventional dark matter scenarios, such sub-keV thermal dark matter would be excluded by free streaming bounds, the constraints are ameliorated in models with dark matter at lower temperature than conventional thermal scenarios, such as the  Flooded Dark Matter model that we have previously considered. Modifying the arguments of Tremaine and Gunn we derive a conservative lower bound on the mass of fermionic dark matter of 70~eV and a stronger lower bound from Lyman $\alpha$ clouds of about 470~eV, leading to slightly smaller cores than have been observed. We comment on this result and how the tension is relaxed in  dark matter scenarios with non-thermal momentum distributions. 

\end{abstract}

\begin{keywords}
Cosmology: theory, dark matter, elementary particles -- galaxies: dwarf.
\vspace{-4mm}
\end{keywords}


\section{Introduction}

The cold dark matter paradigm ($\Lambda$CDM) provides a remarkably good description of cosmology and astrophysics. However,   observations connected with small scales might be in tension with this framework. In particular, structure formation simulations to date assuming $\Lambda$CDM predict a greater number of galaxy satellites and suggest that the density profiles of dwarf spheroidal galaxies (dSphs) should exhibit cusps, in contrast to observations \citep{deBlok:2009sp,Weinberg:2013aya}. Some cores might be explained by baryonic feedback e.g.~\citep{Weinberg:2001gm,Mashchenko:2007jp,Pontzen:2011ty,Nipoti:2014xha}, but this is still undecided \citep{Sellwood:2002vb,Dubinski:2008yi,Jardel:2008bi,Penarrubia:2012bb,Marinacci:2013mha}, indicating the possibility of a more fundamental discrepancy. It is of interest to look at potential resolutions from dark matter effects beyond $\Lambda$CDM, such as warm dark matter \citep{Dalcanton:2000hn}, self-interactions \citep{Spergel:1999mh}, or boson degeneracy \citep{Ji:1994xh,Hu:2000ke,Goodman:2000tg,Peebles:2000yy,Hui:2016ltb}.

An interesting alternative resolution was proposed by \cite{Destri:2012yn} as well as \cite{Domcke:2014kla} (see also \cite{Alexander:2016glq}), who observed that a sub-keV fermion dark matter particle could resolve the core-cusp problem when quantum pressure due to Fermi repulsion provides the dominant support in the central regions of dwarf galaxies. 
In this scenario, the cores of dwarf galaxies are highly analogous to white dwarfs or neutron stars, both of which are supported by degeneracy pressure. This proposal would have the remarkable implication that the structure of dwarf galaxies arises due to a quantum effect on a cosmic scale.

However,  dark matter composed of a  thermal relic lighter than a few keV would erase small scale density perturbations, in conflict with observations \citep{Bode:2000gq,Bolton:2004ge,Viel:2013fqw}. Thus the prospect of Fermi pressure playing a role in the structure of dwarf galaxies is lost for dark matter at the baryonic temperature. However, this scenario might be realized when dark matter is not in thermal contact with baryons so it is cooler --    
when dark matter is decoupled and is not reheated by the Standard Model, for example. The dark matter would then be non-relativistic earlier than a thermal relic, mitigating free streaming bounds. For models that realize such a scenario see for instance: \cite{Feng:2008mu,Berezhiani:1995am}

Another possibility is that the lower limit on the dark matter mass is relaxed by a late-entropy release, as with ``Flooded Dark Matter''  (FDM) \citep{Randall:2015xza}. In this class of models the Standard Model undergoes late-time reheating via entropy injections from long-lived heavy states decaying to the visible sector. We provide a brief outline of the minimal model of Flooded Dark Matter in Appendix \ref{sec:FDM}.
We recast the warm dark matter mass bounds derived from Lyman-$\alpha$ data into constraints on models such as this one with colder temperature and find the following mass limit for Flooded Dark Matter
  \beq
 m_{\rm FDM} \gtrsim  470~{\rm eV} N_f^{-\nicefrac{1}{4}}~,
 \eeq
in terms of $N_f$  `flavors' of Majorana fermion dark matter. This value is borderline for resolving the core-cusp problem but might work well with baryonic feedback in explaining the variability in core sizes that is observed so far. 
The full set of constraints we present is:

\begin{itemize}

  \setlength\itemsep{1.5mm}

\item Lower mass bound from fit to cores: \cite{Domcke:2014kla} find a preference for a narrow dark matter mass range $100$ eV $\lesssim m\lesssim 200$ eV. Once we include the effects of a thermal envelope these constraints reduce to $m \gtrsim 50$ eV, with no upper bound on the mass.

\item Upper mass bound from core size:  Both \cite{Walker:2011zu} and \cite{Amorisco:2012rd} find a  two sigma constraint of $R_c(2\sigma) > 200$ Parsec, which would imply $m\lesssim310$ eV. When we partially redo their analysis with a Burkert profile for dark matter we obtain smaller preferred values $R_c = 0.63^{+0.17}_{-0.13}$ kpc. Our limited analysis does not recover their result  but when we fit their data using a Fermi core with two scales (core radius and scale radius), we obtain an even smaller core $R_c = 0.45^{+0.1}_{-0.1}$ kpc, consistent with a  larger fermion mass of $m\lesssim280$~eV. We can conclude, based on comparing our one-scale and two-scale results, that that the profile with the Fermi-degenerate core (and hence two scales) should reduce the lower bound on the core size and increase the upper bound on the mass of the fermion  to $m< 400$~eV.
\item Lower mass bound from largest allowed core size: \cite{Amorisco:2012rd} also put an upper bound on core size at $2\sigma$ level $R_c(2\sigma) < 2.6$ kpc for the Fornax dSph which implies $m\gtrsim70$~eV lower mass bound. 

\item Lower mass bounds from black hole formation: We show that $m \gtrsim 0.1$ eV suffices to avoid forming black holes out of degenerate Fermi gas cores of dwarf galaxies.

\item  Lyman-$\alpha$: For constant $n_s$ \cite{Baur:2015jsy} found  $m_{\rm WDM}>2.93$ keV, corresponding to $m \gtrsim 530$ eV as we will show in eq.~\ref{eq:wdmfdmeq}. The most conservative, even weaker, bound for warm dark matter was $m_{\rm WDM}> 2.5$ keV, corresponding to $m \gtrsim 470$ eV. We later argue that using a flavored dark matter model we can relax the Lyman-$\alpha$ bounds arbitrarily, at the cost of more complicated dark matter model. Alternatively, while the Lyman-$\alpha$ constraint ($m \gtrsim 470$ eV) is in tension with the mass range for which appropriately large cores are obtained ($m < 400$ eV), this bound can be reconciled if the dark matter has a non-thermal momentum distribution. We will discuss one way to achieve such a nonthermal scenario.

\end{itemize}
 In any case, the coring mechanism based on Fermi degeneracy operates in conjunction with baryonic feedback and it is quite likely that baryonic feedback creates a large core in the Fornax galaxy for example. As a result it may be the case that only \textit{smaller} cores in systems such as Draco need to be resolved by new physics, in which case the Lyman-$\alpha$ constraints become irrelevant.

In what follows,  we extend and improve the analysis of \cite{Domcke:2014kla}.  The earlier study made the simplifying assumption of a completely degenerate Fermi gas, which is not appropriate for realistic scenarios. We treat the dark matter as a quasi-degenerate Fermi gas surrounded by a thermal envelope. Furthermore, we highlight concerns regarding the criteria used in matching to observed data and clarify the condition under which dSphs would be successfully cored through Fermi pressure. 
  
The structure of this paper is as follows: In Section \ref{sec:fermirep} we outline the physics of self gravitating Fermi gases and the role of Fermi repulsion in determining the density profile. We argue that even with sharp power law density profiles dwarf galaxies cannot accommodate arbitrarily large central densities and that Fermi repulsion is generally important in the central region. 
In Section \ref{sec:fits}, using the empirically motivated criterion that dwarf spheroidal galaxies should feature a core of order a few hundred pc, we determine the parameters that yield appropriate sized cores. We point out essential details, such as the importance of stellar anisotropy degeneracy with reference to \cite{Domcke:2014kla}. We also supplement with newer data to reproduce observed properties of the classical Milky Way dwarf galaxies. In  Section \ref{sec:lowermassbounds}, we derive a number of lower bounds on the mass of fermion dark matter. In Section \ref{sec:chandra}, we discuss the lower bound on the mass of fermion dark matter from the requirement that dwarf cores do not collapse into black holes. Appendix \ref{ApB} presents certain definitions and derivations that we use in the main text and Appendix \ref{Apmodels} gives more details about the models we discuss in this paper.


\section{Fermi Repulsion}
\label{sec:fermirep}

We first argue that in the absence of a central black hole,   sufficiently light fermionic dark matter manifests Fermi repulsion that would be relevant in the central region of dwarf galaxies. Assuming a cuspy profile as a starting point, there would always exist a central region with density high enough that the Fermi velocity would exceed the escape velocity so the density in the central region would be inconsistent with the Pauli exclusion principle and would have to be reduced. We later show that full solutions including Fermi pressure lead to cored density distributions.

A fermionic gas of density $\rho$ is degenerate at temperatures below
\beq
T_{\rm Deg}&\sim\frac{h^2}{2\pi m} \left(\frac{\rho}{2m}\right)^{\nicefrac{2}{3}}~,
\label{eq:deg}
\eeq
where $m$ is the mass of the fermion.
In this regime the highest occupation level is $p_F$ (see e.g.~\cite{Landau})
\beq
p_F = m v_F = h \left(\frac{3}{\pi^2 N_f} \frac{\rho}{m}\right)^{\nicefrac{1}{3}}~,
\eeq 
leading to the Fermi pressure in the non-relativistic limit:
\beq
P_{F}=\frac{8\pi}{3h^3}\int_0^{p_F}{\rm d}p\left(\frac{p^4}{\sqrt{p^2+m^2}}\right) = \frac{h^2}{5m^{\nicefrac{8}{3}}}\left(\frac{3}{8 \pi N_f}\right)^{\nicefrac{2}{3}} \rho^{\nicefrac{5}{3}}~,
\label{eq:PF}
\eeq
which at high density and low temperature dominates over the classical pressure (we choose units such that $k=1$)
\beq
\Pcl=\rho\left(\frac{T}{m}\right).
\label{eq:cl}
\eeq
A typical temperature at any point inside this ball is determined by the virial condition
\beq
T(R)=\frac{GM(R)m}{2R}.
\eeq 
Therefore, the classical pressure can be estimated to be
\beq
\Pcl=\frac{GM(R)\rho(R)}{2R}.
\label{eq:cl2}
\eeq
When  $T\gtrsim T_{\rm Deg}$ the classical pressure term dominates.  The  pressure of the gas is then
\begin{equation}
P = 
\begin{cases} 
      \frac{GM(R)\rho(R)}{2R} &\hspace{2mm} T \geq T_{\rm deg} \\
      \frac{h^2}{5m^{\nicefrac{8}{3}}}\left(\frac{3}{8 \pi N_f}\right)^{\nicefrac{2}{3}} \rho(R)^{5/3} &
      \hspace{2mm} T \leq T_{\rm deg}
   \end{cases}~~.
\label{eq:pressure}
\end{equation}

\subsection{Power Law Profiles}
We now argue that systems described by integrable power-laws will generically become quasi-degenerate. It is useful to parameterize the density as
\beq
 \rho=\rho_0\left(\frac{R_0}{R}\right)^{n}~.
\label{den}
 \eeq
Integrable profiles satisfy the condition 
\beq
\frac{{\rm d} \log\rho }{{\rm d} \log R} > -3 \qquad {\rm as}~R\rightarrow 0~.
\label{eqR}
\eeq
We claim that for an integrable density profile one can always find a radius below which the Fermi velocity is larger than escape velocity $v_\infty$, leading to an inconsistent density profile, because states with $v\gtrsim v_\infty$ will escape the gravitational potential, thus reducing the central density. Hence, the system would dynamically relax through evaporative cooling to a quasi-degenerate Fermi gas even from an initially non-degenerate configuration. 

Observe that eq.~(\ref{den}) implies a mass profile of the form
 \beq
 M(R)=4\pi\int_0^R R'{}^2\rho(R')~{\rm d}R' =4\pi\left(\frac{\rho_0 R_0^n }{3-n}\right)R^{3-n}~.
 \label{n<3}
 \eeq
It follows that the virial velocity can be written
\beq
v_{\rm vir}^2
\equiv\frac{GM(R)}{R}\simeq 4\pi G R^{2-n} \left(\frac{\rho_0 R_0^n }{3-n}\right)~.
\label{vir}
\eeq
For the given local density $\rho(R)$, the Fermi surface lies at
\beq
p_F= \hbar \left(\frac{3\pi^2\rho}{m}\right)^{\nicefrac{1}{3}} = \hbar \left(\frac{3\pi^2\rho_0 R_0^n}{m R^n}\right)^{\nicefrac{1}{3}},
\eeq
which determines the Fermi velocity 
\beq
v_F = p_F/m \equiv\frac{\hbar}{m^{\nicefrac{4}{3}}}\left[\frac{3\pi^2 \rho_0 R_0^n}{R^n }\right]^{\nicefrac{1}{3}}.
\eeq
There is some critical radius $R_*$ for which $v_{\rm vir}= v_F$. This roughly coincides with the point at which $T<T_{\rm Deg}$. For an integrable profile ($n<3$)
\beq
R_*^{(n)}=R_0^{n/(6-n)} \left(\frac{\hbar^6}{G^3 \rho_0m^8}\frac{9\pi^3}{4}\right)^{1/(6-n)}~.
\label{rc0}
\eeq
We trade the unknown parameters $\rho_0$ and $R_0$ for the measured mass $M_{\nicefrac{1}{2}}$ within a half-light radius $R_{\nicefrac{1}{2}}$, defined by
\beq
4\pi\left(\frac{\rho_0 R_0^n }{3-n}\right)=  \frac{M_{\nicefrac{1}{2}}}{R_{\nicefrac{1}{2}}^{3-n}}~,
\label{rho-0}
\eeq
in order to obtain
\beq
R_*^{(n)}=\left[\frac{\hbar^6}{m^8}\frac{9\pi^2(3-n)^2}{16}\frac{1}{G^3}\left(\frac{R_{\nicefrac{1}{2}}^{3-n}}{M_{\nicefrac{1}{2}}}\right)\right]^{1/(6-n)}~.
\label{rc1}
\eeq
Degeneracy effects are important in the region $R<R_*$. 
For example for $n=2$:
\beq
R_*^{(2)} \simeq 160~{\rm pc}\left(\frac{300~{\rm eV}}{m}\right)^{2}
\left(\frac{R_{\nicefrac{1}{2}}}{100~{\rm pc}}\right)^{\nicefrac{1}{4}}
\left(\frac{10^8M_\odot}{M_{\nicefrac{1}{2}}}\right)^{\nicefrac{1}{4}}.
\eeq
Similarly for $n=0$:
\beq
R_*^{(0)}=130~{\rm pc}\left(\frac{300~{\rm eV}}{m}\right)^{4/3}
\left(\frac{R_{\nicefrac{1}{2}}}{100~{\rm pc}}\right)^{\nicefrac{1}{2}}
\left(\frac{10^8M_\odot}{M_{\nicefrac{1}{2}}}\right)^{\nicefrac{1}{6}}.
\label{rc2}
\eeq
More generally, for $0\leq n<3$ solving consistently gives a core of $\mathcal{O}(100)$ pc for dark matter masses of a few hundred eV, for typical values of $M_{\nicefrac{1}{2}}$ and $R_{\nicefrac{1}{2}}$.

Having shown that a integrable profile always leads to Fermi degeneracy (for a sufficiently small mass-dependent region), we now consider non-integrable profiles (which do not satisfy the  inequality of eq.~(\ref{eqR})).
 Since these profiles lead to an unphysical infinite total mass in the center of the distribution they must be cut off at some small radius $R=\epsilon$. Below the cutoff $\epsilon$ the profile transitions to an integrable profile. Rather than focus on the outer region, we consider the central cutoff region, which we parameterize solely by the mass $M$ enclosed and  the cut-off radius $\epsilon$. 

We determine the minimum radius required to fit a typical dwarf galaxy mass inside a given size as a function of the fermion mass, where we  require that the escape velocity is larger than the Fermi velocity everywhere inside the gas. This happens for
\beq
v_{\rm esc}^2 = \frac{GM}{\epsilon} > v_F^2 = \frac{\hbar^2}{m^{\nicefrac{8}{3}}}\left(  3\pi^2 \rho \right)^{\nicefrac{2}{3}}~.
\eeq
The least stringent requirement is found when the matter is evenly distributed over the whole region, generating the lowest maximum density, in which case the system must satisfy:
\beq
\frac{GM}{\epsilon} > \frac{\hbar^2}{m^{\nicefrac{8}{3}}}\left(  \frac{9\pi}{4} \frac{M}{\epsilon^3} \right)^{\nicefrac{2}{3}}~.
\eeq
The above inequality implies a lower bound on the cut off
\beq
\epsilon &> \frac{\hbar^2}{G M^{\nicefrac{1}{3}}m^{\nicefrac{8}{3}}} \left( \frac{9\pi}{4}\right)^{\nicefrac{2}{3}} \\
&\gtrsim 180\;\mathrm{pc} \left(\frac{10^8 M_\odot}{M}\right)^{\nicefrac{1}{3}} \left(\frac{300\;\mathrm{eV}}{m}\right)^{\nicefrac{8}{3}}~.
\eeq
Therefore for light fermion masses, $\epsilon$ needs to be similar to, or greater than, the dwarf galaxy size. In other words, the system should be described by an integrable profile over the whole range.
 
 One might suppose that rather than a soft profile below the cutoff of a sharp profile one could encounter a sequence of nested singular profiles. However, ultimately this sequence must be cut off at some $R=\epsilon$ at which the profile becomes non-singular, and the argument above still applies. Having reduced to the previous case, we conclude that a Fermi degenerate region must always be present in a physically relevant region.

We next solve the hydrostatic equilibrium equations explicitly to derive the density profile accounting for Fermi degeneracy to show that  the result is a cored central region.


\subsection{Static Solutions}

We are interested in determining the density profile of a self-gravitating ball of fermions. Such a static density profile has to satisfy three equations:\\[3pt] {\em i)} hydrostatic equilibrium
\beq
\frac{{\rm d}P}{{\rm d}R} &=-\left(\frac{GM(R)}{R^2}\right)\rho(R)~,
\label{eq:hydroeq1}
\eeq
{\em ii)} pressure equality
\beq
P(R) & = P_{\rm cl}+ P_F \\
& \approx \frac{GM(R)\rho(R)}{2R} + \frac{h^2}{5m^{\nicefrac{8}{3}}}\left(\frac{3}{8 \pi N_f}\right)^{\nicefrac{2}{3}} \rho(R)^{\nicefrac{5}{3}}~, \label{eq:hydroeq2}
\eeq
 {\em iii)} the continuity condition
\beq
M(R) &= 4\pi \int_0^R \rho(r) r^2 {\rm d}r~.
\label{eq:hydroeq3}
\eeq
Equations (\ref{eq:hydroeq1})-(\ref{eq:hydroeq3}) with initial conditions $\rho(0) = \rho_0$, and $M(0) = 0$ form a closed system that can be solved to obtain a self-consistent static density profile of a gas of fermions. 

First we construct solutions supported by degeneracy pressure alone, by setting $\Pcl = 0$. In this case eqs.~(\ref{eq:hydroeq1}), (\ref{eq:hydroeq2}) \& (\ref{eq:hydroeq3}) reduce to
\beq
\frac{h^2}{3m^{\nicefrac{8}{3}}}\left(\frac{3}{8 \pi}\right)^{\nicefrac{2}{3}} \frac{\rm d}{{\rm d}R} \left( \frac{R^2}{\rho(R)^{\nicefrac{1}{3}}} \frac{{\rm d} \rho(R)}{{\rm d}R}\right) = -4\pi G R^2 \rho(R).
\label{eq:fermialone}
\eeq
Note that for small $R$ an approximate solution is given by $\rho = \rho_0 [1-(R/R_a)^2]$, for $R_a$ an appropriately chosen scale
\beq
R_a^2 = \frac{h^2}{2\pi G m^{\nicefrac{8}{3}}\rho_0^{\nicefrac{1}{3}}}\left(\frac{3}{8 \pi N_f}\right)^{\nicefrac{2}{3}}~,
\label{Ra}
\eeq
in agreement with the approximate analysis in the previous section. We can also see that this scale arise in eq.~(\ref{rc0}) for the choice $n=0$. 
This indicates we should expect a constant density core of size $R_a$ as Fermi pressure flattens the expected cusps of density distributions of dwarf galaxies.

 \begin{figure}
\includegraphics[width=0.45\textwidth]{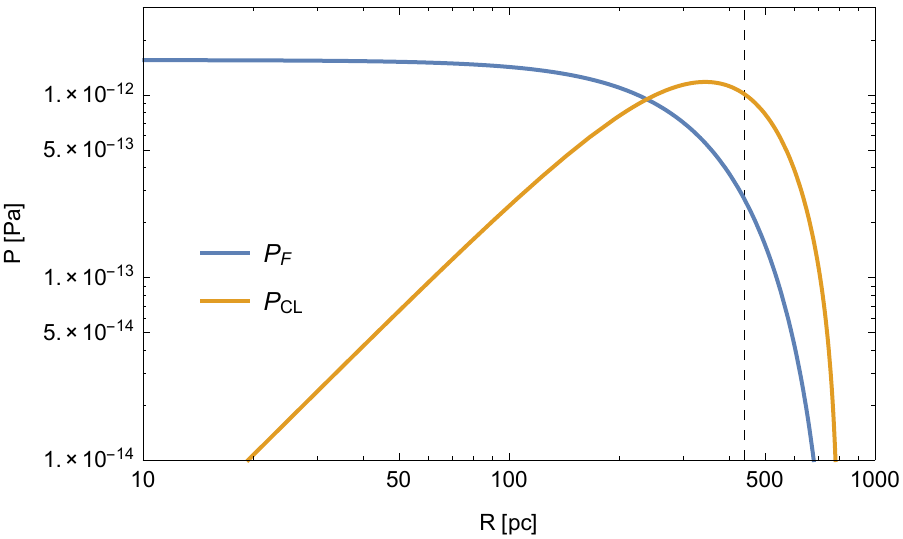}
\caption{
The behavior of the classical ({\sc yellow}) and quantum ({\sc blue}) pressure terms as a function of radial distance, where we have taken dark matter mass of $m=200$ eV and central density $\rho_0=10^{-20}~{\rm kg/m}^3$. This implies $R_a\sim440$pc (vertical line), the critical radius at which the associated  fully degenerate profile terminates, as given in eq.~(\ref{eq:scalingr0}). Note that at $R\sim R_a$ the contribution of each pressure component is significant.
\label{fig:pclpf}}
\end{figure}

We now look at the full solution. Eq.~(\ref{eq:fermialone}) is an example of a Lane-Emden equation with index $n=\nicefrac{3}{2}$. The solutions to this differential equation are polytropes with index $\gamma = \nicefrac{5}{3}$, see e.g.~\citep{Jaffe:2006,Domcke:2014kla}. Indeed, these solutions are constant as $R \rightarrow 0$, acting as cored profiles. Moreover, the density profile of a fully degenerate Fermi gas has finite extent with density vanishing at
\beq
R_0^2 = \xi_1^2 \frac{h^2}{8\pi G m^{\nicefrac{8}{3}} \rho_0^{\nicefrac{1}{3}}}\left(\frac{3}{8 \pi N_f}\right)^{\nicefrac{2}{3}},
\label{eq:scalingr0}
\eeq
where $\xi_1 = 3.65$ is a numerical constant. 

Figure~\ref{fig:pclpf} illustrates the behavior of the two pressure components for an example with dark matter mass of $m=200$ eV and central density $\rho_0= 10^{-20}~{\rm kg/m}^3$. With these parameter values, $R_a\simeq 440~{\rm pc}$ (taking $N_f=1$). We see that close to $R_a$, the classical pressure is as important as the Fermi pressure.

\begin{figure}
\includegraphics[width=0.45\textwidth]{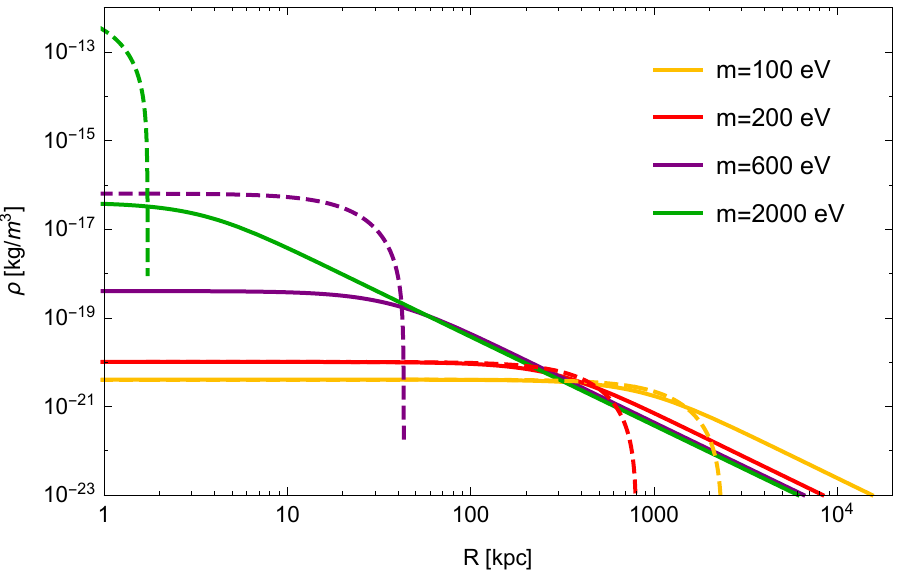}
\caption{
We treat the dark matter as a quasi-degenerate self-gravitating Fermi gas for constituent fermions of various masses and different central densities (as can be read off from the plot).  We take the central densities of these profiles such that they reproduce the $M_{\nicefrac{1}{2}}$ for the Fornax dSph. Observe for 2 keV fermions the  profile is decidedly cuspy, whereas the profiles for 100 eV and 200 eV dark matter exhibits flat cores of a few hundred parsecs. The dashed curves correspond to solutions to eq.~(\ref{eq:fermialone}), and give the distributions corresponding to fully degenerate Fermi gases.
\label{fig:rhovsr}}
\end{figure}

To increase the range of validity of our density profiles we investigate the full solutions to eq.~(\ref{eq:hydroeq1})--(\ref{eq:hydroeq3}) including both $\Pcl$ and $P_F$.  Figure~\ref{fig:rhovsr} shows a set of solutions to these equations including both classical and degeneracy pressure (we fix the central densities of these solutions such that they reproduce the $M_{\nicefrac{1}{2}}$ for the Fornax dSph). These solutions are constant for $R\rightarrow 0$ just as those supported by the Fermi pressure alone. However, the full solutions do not vanish at some finite $R$. In fact  $\rho \sim R^{-2}$ is a solution for $R\rightarrow \infty$ as the classical pressure term takes over.

There is a characteristic length scale associated with the transition from constant density core to isothermal behavior. We can define this scale as the radius $R_P$ at which the $\Pcl = P_F$. 
We choose to define the core radius $R_c$ as a radius at which the slope of the density distribution reaches a particular value and we use the definition of~\cite{Burkert:2015vla}:
\beq
\left.\frac{{\rm d\; log}\;\rho}{{\rm d\; log}\;R}\right|_{R_c}=-\frac{3}{2}~.
\label{32}
\eeq
This definition allows us to compare our results with the results of others in a model-independent way. Notice that $R_0 \sim R_a \sim R_*^{(0)} \sim R_P$, and similarly $R_c \sim R_0$. This is not surprising as there is only one relevant combination\footnote{It is in fact possible to construct additional length scales, however, these are many orders of magnitude smaller or larger and hence irrelevant.}  of $G,h,m$ and $\rho_0$ with dimensions of length namely eq.~(\ref{eq:scalingr0}).

\section{Fits to measured dwarf galaxies}
\label{sec:fits}

In the previous section we have shown that including Fermi pressure in the hydrostatic equations for a self-gravitating gas of light fermions flattens their density profiles to constant density cores with mass-dependent characteristic sizes. We also showed that including the classical pressure term extends otherwise finite extent solutions to include isothermal tails. In this section we fit these quasi-degenerate profiles to the kinematic properties of the eight classical dwarf galaxies: Carina, Draco, Fornax, Leo I, Leo II, Sculptor, Sextans, and Ursa Minor, see e.g.~\cite{Walker:2009zp}. We first consider the kinematic variables studied in \cite{Domcke:2014kla} and demonstrate that due to a degeneracy this method does not suffice to determine the core sizes.  

\vspace{-3mm}
\subsection{Profile Degeneracy}

\begin{figure}
\begin{center}
\includegraphics[width=0.45\textwidth]{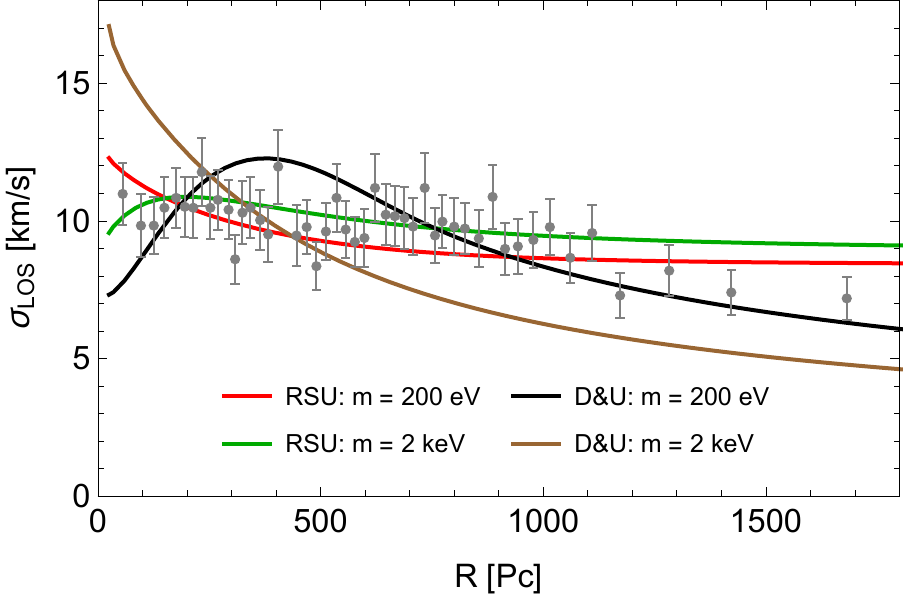}
\caption{We plot data for the $\sigma_{\sc LOS}$ of Fornax as reported in \citep{Walker:2009zp}. The {\sc red} curve is the predicted $\sigma_{\sc LOS}$ for a cored density profile ($R_c\sim400$ pc)  due to 200 eV fermion dark matter, corresponding to our hydrostatic solution (RSU) with Fermi gas. In {\sc green} is a cuspy profile ($R_c\sim10$ pc) assuming 2 keV fermion dark matter. Alongside, in {\sc black}, we display the result of \citep{Domcke:2014kla} for 200 eV dark matter which assumes the Fermi gas is fully degenerate (D\&U) and in {\sc brown} we show a fully degenerate gas with dark matter mass 2 keV, which is very cuspy.}
\label{fig:sigmalos}
\end{center}
\end{figure}

One of the main observables that is used to obtain kinematic information is the projected velocity dispersion along the line-of-sight (LOS) $\sigma_{\sc LOS}$, as defined  in \citep{Binney}:
\beq
\sigma_{\sc LOS}^2(R)=\frac{2G}{I(R)}\int_R^\infty
\nu(R')M(R')(R')^{2\beta-2}F(\beta,R,R')~{\rm d}R',
\label{slos}\eeq
where  $I(r)$ is the projected stellar density, and $\nu(r')$ is the associated 3D stellar density. Following e.g.~\citep{Walker:2009zp}, we use the Plummer profile \citep{Plummer:1911zza}, and we give the forms of $I(r)$, $\nu(r')$ and $F(\beta,r,r')$ in Appendix \ref{ApB}. As an example, we show the data reported in \citep{Walker:2009zp} for the classical dwarf galaxy Fornax in Figure \ref{fig:sigmalos}.

Importantly, fits to $\sigma_{\sc LOS}$ exhibit a well known degeneracy between values of $\beta$ -- the stellar anisotropy -- and deformations in the density profile. The dependence of $\sigma_{\sc LOS}$ on the stellar orbital anisotropy $\beta$ arises because we can observe only the line of sight velocities which forces us to project the two different velocity components $\{v_R,v_\theta\}$ onto the line of sight:
\beq
v_{\sc LOS}=v_R \cos\theta-v_\theta\sin\theta~.
\eeq 
In order to recover the lost information one typically assumes there is some overall relationship between $v_R$ and $v_\theta$ parametrized by $\beta$ as
\beq 
\beta=1-\frac{\langle v_\theta^2\rangle}{\langle v_R^2\rangle}~.
\eeq

Figure~\ref{fig:sigmalos} gives the observed $\sigma_{\sc LOS}$ for Fornax, and we can compare this to the prediction for a cored profile (e.g.~200 eV fermion dark matter) and also to a decidedly cuspy profile (e.g.~2 keV fermion dark matter). As can be seen both are in good agreement with the data and so $\sigma_{\sc LOS}$ does not discriminate between cored and cuspy profiles. 

This issue is not apparent in \cite{Domcke:2014kla} because the analysis considers only fully degenerate Fermi distributions. The finite-extent step-like density profile of a degenerate Fermi-gas would predict decreasing $\sigma_{\sc LOS}$ outside the core region. Since the data indicates a nearly constant $\sigma_{\sc LOS}$ over the entire observed dwarf, the fully degenerate model is best fit with large core radii. This gives the misleading impression that the core-cusp problem is manifested in the $\sigma_{\sc LOS}$  data as can be seen in the $\chi^2$ analysis in Figure \ref{fig:chisqvsm}. Once the thermal tail is included, the predicted $\sigma_{\sc LOS}$ no longer drops outside of the core region ($1/R^2$ profiles have flat rotation curves) and the cores are allowed to be nearly arbitrarily small (or equivalently the profile arbitrarily cuspy) by the data. We conclude that  the observed $\sigma_{\sc LOS}$ can be fit equally well by cuspy and cored profiles by adjusting $\beta$ \citep{Evans:2008ik} in the range $\beta \in (-2,1)$.
 
 In Figure~\ref{fig:chisqvsm} we display in blue the $\chi^2$ per degree of freedom for fits to the eight classical dwarf galaxies as the dark matter mass is varied, marginalizing over $\beta$ and $\rho_0$. Our fits suggest that dark matter with mass $m>100\;\mathrm{eV}$ matches the kinematic data very well. The flatness of the $\chi^2$ above $m>100$ eV in Figure \ref{fig:chisqvsm} indicates that the data does not definitely establish existence of cores due to the $\beta$ degeneracy inherent in fitting to $\sigma_{\sc LOS}$ data.

In Figure \ref{fig:rhocvsrc} we show the fitted core densities and core sizes of the eight classical dwarfs that arise as our best fit parameters for different dark matter masses from the $\sigma_{\sc LOS}$ data of ~\citep{Walker:2009zp}.

\vspace{-3mm}
\subsection{Beyond $\beta$-Degeneracy: Core Sizes}
\label{sec:beyondbeta}

New approaches circumvent the  $\beta$-degeneracy \citep{Battaglia:2008jz,Walker:2011zu,Agnello:2012uc,Amorisco:2012rd} and can be used to more robustly determine core sizes of dwarf spheroidals in which star subpopulations can be separated. Using these more recent results we show that there is a range of fermion masses with cores of the right order of magnitude and which fit the stellar kinematics. The approach of \cite{Battaglia:2008jz,Walker:2011zu,Agnello:2012uc,Amorisco:2012rd} is to split the tracer star population chemically into subpopulations and extract the mass enclosed at radii at which the uncertainties due to the stellar orbital anisotropy are the smallest for each subpopulation. The resulting mass profile is only known at two or three radial points, since the subpopulations have different spatial extent, but is nearly insensitive to $\beta$. Studies utilizing this approach have confirmed the presence of flat cores in Fornax and Sculptor and are in broad agreement \citep{Battaglia:2008jz,Walker:2011zu,Agnello:2012uc,Amorisco:2012rd}.

\cite{Walker:2011zu} consider a cored and a cuspy density profile to argue\footnote{M.~G.~Walker, private communication.} that the dSphs Fornax and Sculptor require core radii of order a few hundred pc at $\sim$95\% CL. To match observations we will therefore require a constant density core of magnitude
\beq
R_c\sim{\rm few}\times100~{\rm pc}~.
\eeq
Similarly \cite{Amorisco:2012rd} find that at $2\sigma$ confidence level $R_c > 200~{\rm pc}$ for the Fornax dSph.

\begin{figure}
\begin{center}
\vspace{-4mm}
\includegraphics[width=0.4\textwidth]{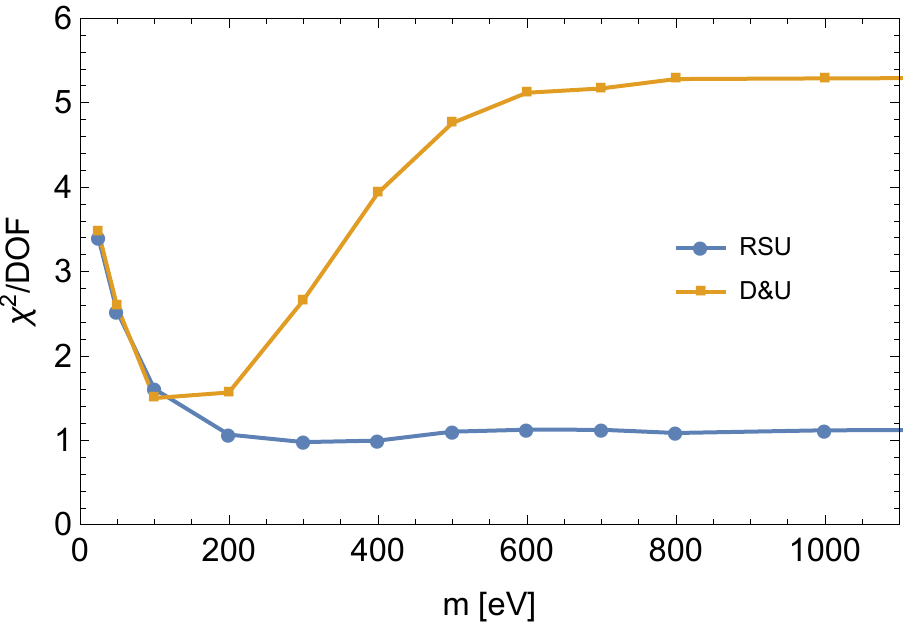}
\caption{The $\chi^2$ per degree of freedom for fits to the eight classical dwarf galaxies as a function of dark matter mass $m$. In {\sc blue} ({\sc yellow}) is the $\chi^2$ associated to fits for a quasi (fully) degenerate Fermi gas. The flat $\chi^2$ exhibited by the {\sc blue} curve for $m>200$ eV  is a result of the $\beta$-degeneracy.
\label{fig:chisqvsm}}
\includegraphics[width=0.4\textwidth]{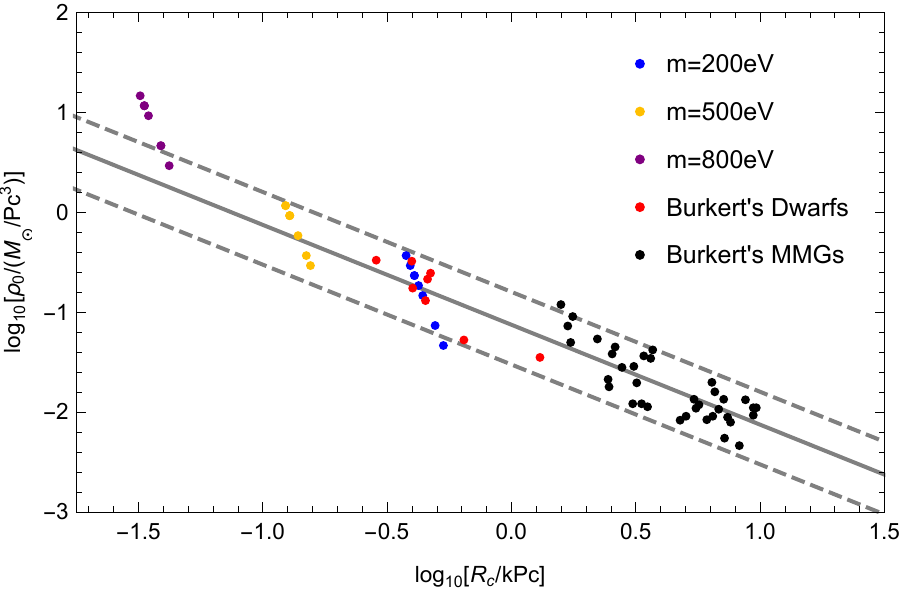}
\caption{
Correlation between core size $R_c$ and central density $\rho_0$ of the eight classical Milky Way dwarf spheroidal galaxies from different fits. ``Burkert's Dwarfs'' ({\sc red}) and ``Burkert's MMGs'' ({\sc black}) refers to fits identified in \citep{Burkert:2015vla} for dwarfs and more massive galaxies, respectively. 
The {\sc blue/yellow/green} points provide fits for the eight classical dwarfs assuming a quasi-degenerate Fermi gas for different dark matter masses. Note that some dwarfs have similar parameter fits and so all eight points are not always distinct.
\citep{Burkert:2015vla} observed that the populations appear to obey the scaling relationship $\langle\rho_0 R_c\rangle= 75^{+85}_{-45}~M_{\odot} ~{\rm pc}^{-2}$, this is indicated by the  {\sc solid}/{\sc dashed} lines. 
\label{fig:rhocvsrc}}
\includegraphics[width=0.42\textwidth]{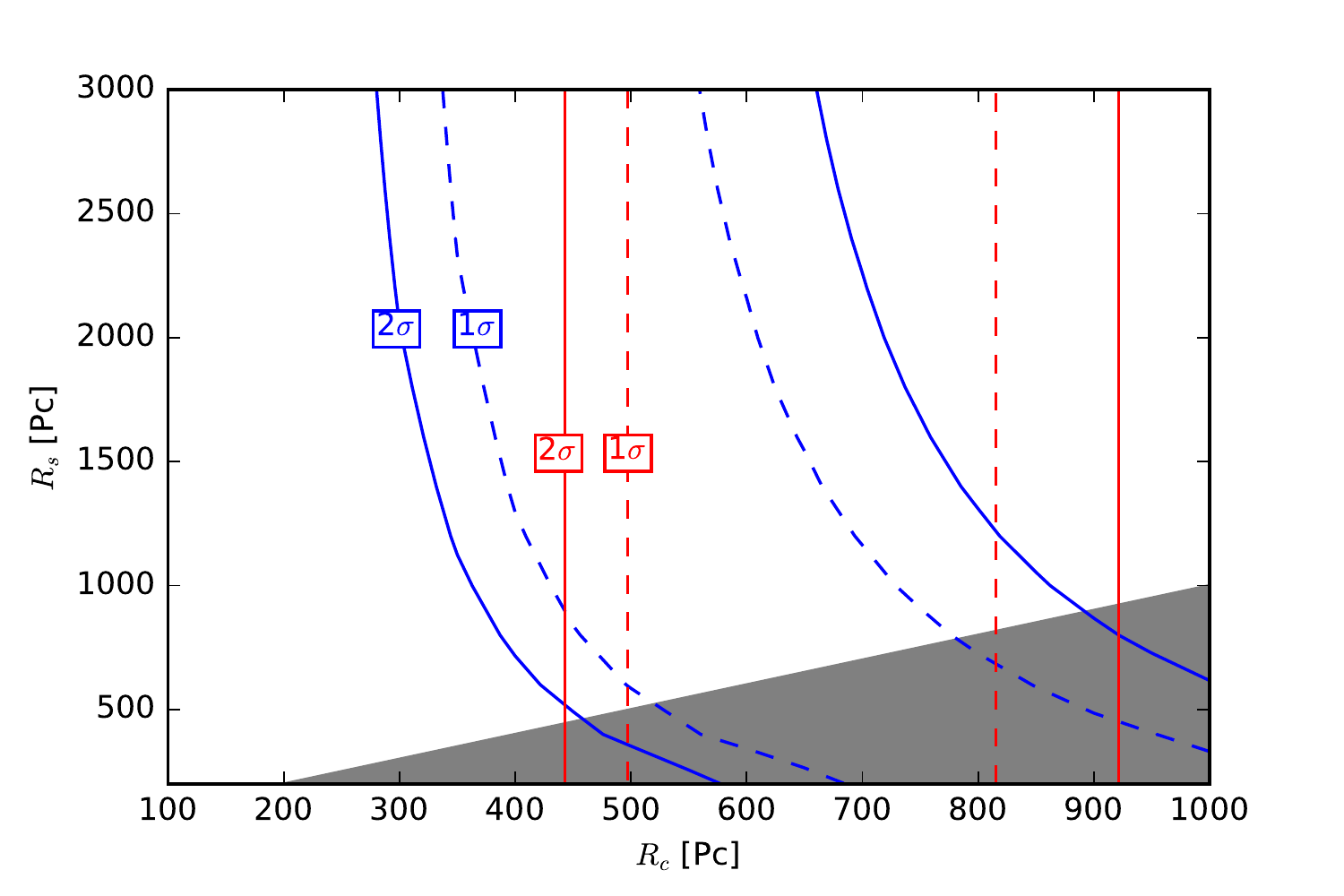}
\caption{ 
Contours of  $\chi^2$ for fits to data of Amorisco et al. The {\sc red} bands bounded by the {\sc dashed} ({\sc solid}) contours show the 1$\sigma$ (2$\sigma$) fits to with a Burkert profile with a single scale parameter $R_c$ (and is therefore independent of $R_s$). The {\sc blue} bands correspond to a fit with a profile from eq.~(\ref{eq:twosc}). The {\sc grey} area corresponds to parameter space $R_c > R_s$. Although not in principle invalid, these solutions do not correspond to quasi-degenerate fermi gas.
\label{fig:twoscale}}
\end{center}
\end{figure}

However, both \cite{Walker:2009zp} and \cite{Amorisco:2012rd} considered profiles that have a steep drop-off in density  ($1/r^3$) outside their core. However, the solution to the hydrostatic equilibrium equation that we derived has an isothermal envelope that behaves like $1/r^2$ (with presumably a steeper drop-off further out). Since both Burkert and NFW profiles use only one scale radius, fitting the kinematic properties of the outer parts of the galaxy makes the core radii of \citep{Walker:2011zu,Agnello:2012uc,Amorisco:2012rd} larger than those obtained using an isothermal profile. The fit is sensitive to both the physics responsible for the core and the physics responsible for the asymptotic behavior of the dwarf galaxy. With only one variable to fit, the derived core radii are some sort of geometrical averages of these two scales. 

Because our profile includes an explicit core, our solution requires that we separate the two scales. In order to make this clear we fit the $\sigma_{LOS}$ data for three distinct stellar subpopulations from \cite{Amorisco:2012rd} to a two scale profile (with one more scale parameter than NFW or Burkert profiles) of the form:
\beq
\rho_{2}(r) = \frac{\rho_0}{\left[1+\left(r/R_c\right)^2\right]\left[1+r/R_s\right]}~.
\label{eq:twosc}
\eeq
This profile behaves like a cored profile up to $r \sim R_c$, transitions into an isothermal profile and then asymptotes to a $r^{-3}$ profile when $r > R_s$. We used the velocity dispersions of three sub-populations of the Fornax galaxy from figure 1 of \cite{Amorisco:2012rd} to constrain a two scale dark matter density profile in Fornax.

We fit the data in a following way: for each combination of $R_c, R_s$ we sample the four dimensional space of proposed central density $\rho_0$ of the dark matter density profile and three independent velocity anisotropies of each star sub-population $\beta_i$. We marginalized over this four-dimensional parameter space (central density, three independent $\beta_i$ for each sub-population) to obtain the best $\chi^2$ for a given $R_c,R_s$ combination. Our fit indicates that allowing for a second scale relaxes the lower bound on the core radius to below 300 pc as can be seen in Figure~\ref{fig:twoscale}.

However, when we redo the Amorisco~\textit{et~al.} analysis for a Burkert profile, we do not reproduce their results: our fit of the Burkert profile leads to $R_c = 650\pm130$ pc instead of $R_c = 1000\pm400$ pc from \cite{Amorisco:2012rd}. We are unable to resolve this discrepancy without a careful reanalysis of the original data. Our incomplete reanalysis of Amorisco~\textit{et~al.}  would lead to a stronger bound which is however less reliable than the original bound. However, based on comparing our one-scale and two-scale results,  the two-scale profile tends to reduce the lower bound on core size compared to the Burkert profile by about a factor of $\sim1.5$ cf. Figure~\ref{fig:twoscale}. Applying this factor, we assume that a full analysis with a two-scale profile would lead to reduced lower bound on the core size somewhere in the proximity of $R_c > 200/1.5$~pc~$\sim 130$~pc. This lower bound on core size would correspond to a reduced upper bound of $m<400$~eV.

We illustrate these results in Figure \ref{fig:fornax}. This plot shows our prediction for the core radius in the Fornax dSph as function of $m$, by demanding that the profile reproduces the observed half-light mass $M_{1/2}$ as given in \cite{Walker:2009zp}. The blue band indicates the core size preferred by \cite{Amorisco:2012rd} ($ 200~{\rm pc} < R_c < 2600~{\rm pc} $).  Should these be relaxed by a two scale fit, we also mark a weaker condition $R_c > 130~{\rm pc}$ by a green band.
 Our fits to the core radius data of \cite{Amorisco:2012rd} favor masses of 70 eV -- 320 eV and extend to 400 eV once we consider the effect of a two-scale fit.

Some variation in cores size is expected due to differences between dwarfs, e.g.~Fornax has stellar mass $8\times10^6~M_\odot$ \citep{deBoer:2012py}, while Sculptor has $3\times10^7~M_\odot$ \citep{deBoer:2012dv}. Variations in the core size among dwarfs in this model is a result of differences in their central densities.


\section{Lower Mass Bounds}
\label{sec:lowermassbounds}

Although Fermi repulsion might improve the density profiles of light fermionic dark matter, too low a mass would lead to overly large cored profiles that would disagree with observed structures of dwarf galaxies. Too light dark matter would also prevent the formation of structure on small mass scales, leading to disagreement with Lyman-$\alpha$ constraints on the matter power spectrum. This latter bound depends on the relative temperature of the dark matter, which leads to model dependence. We now derive both bounds.

\subsection{Dwarf Galaxy Lower Mass Constraints}
We focus on the dwarf galaxy Fornax, for which  the core radius  has been measured (albeit with sizable uncertainty) by \cite{Amorisco:2012rd}:
\beq
R_c^{\rm Fornax} = 1^{+0.8}_{-0.4}~{\rm kpc}~.
 \eeq
Fermion masses that would produce cores in excess of $R_c\sim 2.6$ kpc are ruled out at 95\% CL by this measurement.

 \begin{figure}
\begin{center}
  \includegraphics[width=0.45\textwidth]{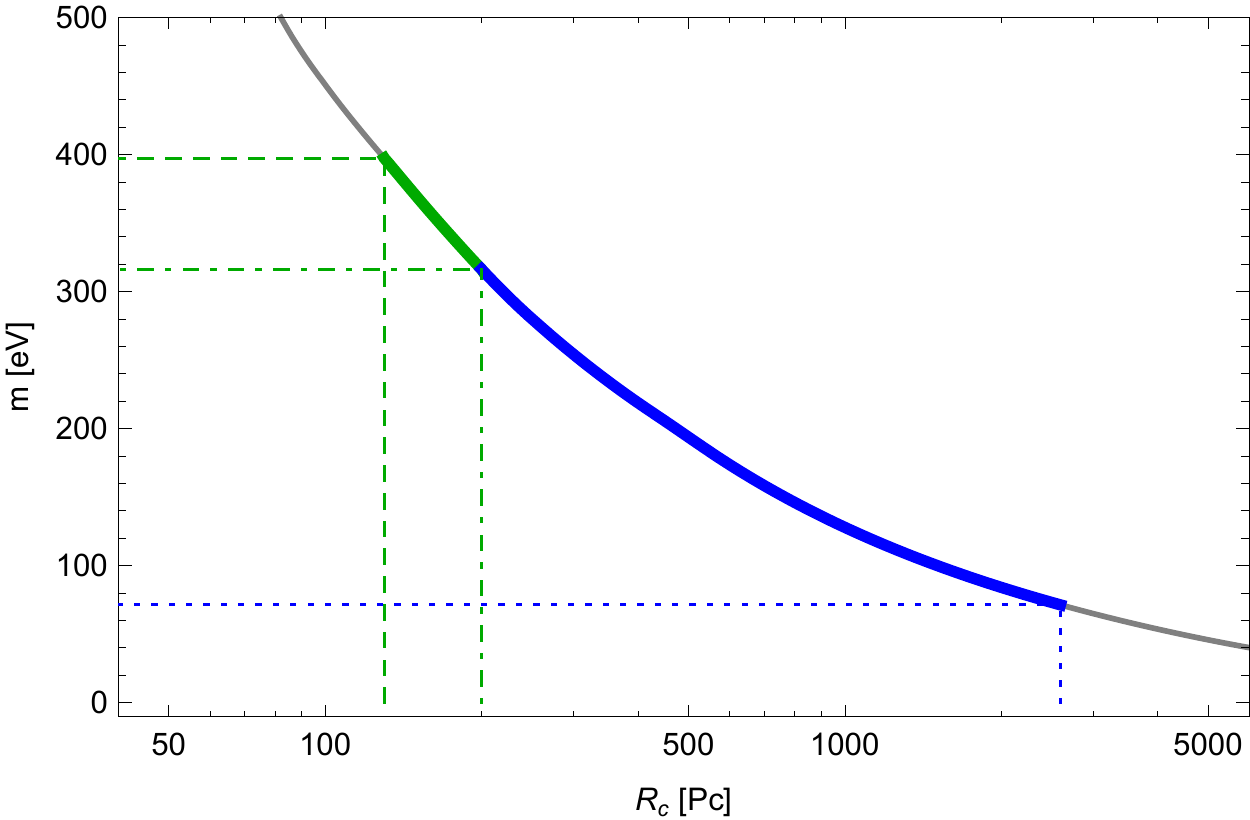}
\caption{Applying the static solution of Section \ref{sec:fermirep} to the Fornax dwarf galaxy, we show the necessary dark matter mass $m$ as a function of the desired core radius $R_c$. The {\sc blue} part of the curve indicates the preferred set of $R_c$ from \protect\cite{Walker:2011zu}. The {\sc green} portion of the curve shows a weaker condition $R_c> 130~\rm{pc}$ .\label{fig:fornax}}
 \end{center}
\end{figure}

If the density profile remains flat to large radius, the central density of Fornax can be self-consistently determined from the half-light radius $R_{\nicefrac{1}{2}}=668\pm34$~pc
 (note that $R_{\nicefrac{1}{2}}<R_c$) and the mass this encloses is  $M_{\nicefrac{1}{2}}=(5.3  \pm0.9)\times10^7M_\odot$ \citep{Walker:2009zp}. Taking the central values gives
\beq
\rho_0=\frac{M_{\nicefrac{1}{2}}}{\frac{4\pi}{3} R_{\nicefrac{1}{2}}^3} \approx3\times10^{-21}~{\rm kg/m}^3.
\eeq
With this central density, this scenario predicts a core in excess of  2.6 kpc for dark matter mass $m\lesssim 70~{\rm eV}$, cf.~Fig~\ref{fig:fornax}. As a result we obtain a bound
\beq
m\gtrsim 70~{\rm eV}~.
\eeq 
This argument is similar to the reasoning underlying the Tremaine-Gunn bound \citep{Tremaine:1979we}, in which the authors used an isothermal profile in the Milky Way galaxy and other structures to put a lower mass bound on dark matter based solely on phase space density considerations, and a slightly weaker bound based on Fermi repulsion.
They too obtained their bound by observing that with too small a mass for the dark matter particle, objects of a given total mass would exceed their observed sizes (this statement can be re-expressed in terms of velocity dispersions).
Here we use the cored profile given by Fermi repulsion and find that the radius of Fornax would exceed its measured value unless a minimum mass bound applies.
Future observations should provide a better understanding of  $R_c$, $R_{\nicefrac{1}{2}}$ and $M_{\nicefrac{1}{2}}$ in Fornax and other dwarf spheroidals, which will in turn improve the lower mass bound for fermion dark matter.

\subsection{Lyman-$\alpha$ bound on light dark matter}

Small scale structure growth also gives  lower bounds on the dark matter mass. The growth of over-densities in the matter distribution is suppressed when dark matter is relativistic. As a result, if dark matter is too light and relativistic during late cosmic evolution, small scale structure over-densities (the matter power spectrum for large $k$) would be suppressed compared to the standard $\Lambda$CDM scenario. But we know that the matter power spectrum  $P(k)$ is not suppressed at large $k \sim \textrm{Mpc}^{-1}$ because we observe absorption lines in the corresponding spectra of quasars due to small hydrogen clouds in their lines of sight -- the so called Lyman-$\alpha$ forests.

Existence of these lines has been interpreted as a lower bound on the mass of dark matter by studies such as \cite{Bolton:2004ge,Viel:2013fqw}. The current bounds on dark matter mass comes from \cite{Baur:2015jsy} who require that conventional thermal dark matter has to have\footnote{Using only the Lyman-$\alpha$ and $H$ constraints \cite{Baur:2015jsy} obtain a limit $m_{\rm WDM}>4.09$ keV. After including additional datasets (Lyman-$\alpha$ + Planck (TT + lowP+ TE + EE) + BAO) they set two further bounds, depending on whether running of $n_s$ is incorporated or not, these are $m_{\rm WDM} >4.12$ keV and $m_{\rm WDM} >2.93$ keV, receptively. Here we adopt the most conservative bound from \cite{Baur:2015jsy} $m_{\rm WDM} \gtrsim 2.5$ keV} $m_{\rm WDM} > 2.5$ keV at $95\%$ CL.

We note that such bounds depend on the dark matter temperature, which is critical to determining which states are relativistic when the Lyman-$\alpha$ producing clouds are formed.  Allowing for dark matter that is colder than the Standard Model sector and  that becomes non-relativistic earlier softens the bounds of \cite{Baur:2015jsy}. Dark matter that freezes out relativistically  can have temperature lower than that of the Standard Model, such as with Flooded Dark Matter \citep{Randall:2015xza}, which we describe in Appendix~\ref{sec:FDM}. 

Here we  give a bound on any dark matter sector, assuming only that the sector is thermal with temperature sufficiently high to allow for the correct energy density. Even weaker bounds can apply for dark matter with a nonthermal distribution, as we discuss later in this section.

In order to map the bounds on Warm Dark Matter onto bounds on models such as Flooded Dark Matter we need to satisfy two conditions. First we require that the Flooded Dark Matter candidate reproduces the observed relic density. This fixes a relationship between the dark matter temperature $T_\DM$, mass $m$ and degrees of freedom $g_\text{DM}$ as indicated in Appendix~\ref{sec:FDM}:
\beq
\frac{T_{\DM}}{T_{\SM}} = \left( \frac{g_{\gamma}}{g_{\rm DM}}  \eta
\frac{\Omega_{\rm DM}}{\Omega_{B}} \frac{m_N}{m}\right)^{1/3},
\label{eq:tdm}
\eeq
with $m_N$ the nucleon mass, and $\eta = n_B/s$. A second condition states that two dark matter models will have the same effect on the power spectrum if their effective number of relativistic degrees of freedom is the same. Once we find a combination of mass $m$ and number of degrees of freedom $g_\text{DM}$ such that we get the correct relic density and effective number of relativistic degrees of freedom identical to that of warm dark matter with mass $m_\text{WDM}$, we can determine whether it is ruled out by \cite{Baur:2015jsy}.

The effective number of relativistic degrees of freedom at a given $z$ is
\begin{equation}
\Delta N_{\rm eff}(z) = \frac{3\langle P(z)\rangle }{\rho_\nu(z)}~, 
\end{equation}
where $\rho_\nu(z)$ is the energy density of neutrinos and $\langle P(z) \rangle$ is the pressure of the dark matter component defined by
\begin{equation}
P(z) = \int \frac{p^2 {\rm d}p}{2\pi} f(p,z) \frac{p^2}{3\sqrt{m^2+p^2}}~.
\end{equation}
To translate the bound of \cite{Baur:2015jsy}, we compare the contribution to the number of relativistic degrees of freedom from a standard warm dark matter model with mass $m_{\rm WDM}$ at temperature $T_{\rm WDM} = T_{\rm SM}$ and another dark matter candidate with mass $m$ and temperature $T_{\rm DM} < T_{\rm SM}$ given by Eq.~(\ref{eq:tdm}). Because each model is described by two scales (temperature and mass) and the overall density is fixed by $\Omega_{\rm DM}$, the ratio of these scales fully characterizes these models. As long as these ratios are equal, the models have the same $\Delta N_{\rm eff}$. The equality
\begin{equation}
\frac{m_{\rm WDM}}{T_{\rm WDM}} = \frac{m}{T_{\rm DM}}~,
\label{eq:eq}
\end{equation}
together with the expression for $T_{\rm DM}$ from eq.~(\ref{eq:tdm}) allows us to establish a correspondence between thermal dark matter and a dark matter model with a temperature given by (\ref{eq:tdm}). Using equations (\ref{eq:tdm}), (\ref{eq:eq}) and that $T_{\rm WDM}=T_{\rm SM}$ we can show that to get an equivalent effect on Lyman-$\alpha$ power spectrum from our model would require
\beq
m &= \left(m_{\rm WDM}^3 m_{\rm B}\right)^{\nicefrac{1}{4}} \left(\frac{g_\gamma}{g_{\rm DM}}\eta\frac{\Omega_{\rm DM}}{\Omega_{B}}\right)^{\nicefrac{1}{4}}  \\
                & = 235\;\mathrm{eV} \left( \frac{m_{\rm WDM}}{1\;\mathrm{keV}}\right)^{\nicefrac{3}{4}} \left( \frac{2}{g_{\rm DM}}\right)^{\nicefrac{1}{4}}~.
\label{eq:wdmfdmeq}
\eeq
We have verified the statement of eq.~(\ref{eq:wdmfdmeq}) by computing the matter power spectra for both models using the CLASS software \citep{Blas:2011rf} for a range of equivalent masses. As a result, the $2.5$ keV lower mass bound for conventional thermal dark matter \cite{Baur:2015jsy} is weakened:
\beq
m_{\DM} > 470\; {\rm eV} \; \left(\frac{2}{g_{\DM}}\right)^{\nicefrac{1}{4}}~.
\label{eq:eqgdm}
\eeq
Observe that the Lyman-$\alpha$ limit is even weaker for larger $g_{\DM}$, i.e. for dark matter with more flavors. Table \ref{tab:masses} shows the explicit bounds as a function of number of flavors of $N_f \equiv g_{\DM}/2$ and for two different bounds on warm dark matter from  \cite{Baur:2015jsy}: the middle column shows a bound assuming $m_{\rm WDM} < 2.5\keV$, the right column shows a bound for $m_{\rm WDM}<2.93\keV$.

Note that the fits of \cite{Evans:2008ik} and \cite{Amorisco:2012rd} favor masses of 70 eV -- 320 eV (400 eV in case we include the two-scale effect). This mass range is in tension with the Lyman-$\alpha$ bound, and thus resolving the core-cusp problem might require additional effects, such as baryonic feedback. There is, however, a caveat to this conclusion, as we discuss next.
\begin{table}
\begin{center}
\begin{tabular}{c|c|c|c}
$N_f$ & Min.~$m_{1,\rm DM}$ & Min.~$m_{2,\rm DM}$ \\\hline
1 & 470 eV & 530 eV\\
4 & 330 eV & 375 eV \\
16 & 235 eV & 265 eV
\end{tabular}
\caption{Minimum FDM mass as a function of number of flavors of mass degenerate Majorana fermions. The middle column shows a bound assuming $m_{\rm WDM} < 2.5\keV$, the right column shows bound for $m_{\rm WDM}<2.93\keV$}
\label{tab:masses} 
\end{center}
\vspace{-2mm}
\end{table}

\subsection{Skewed Dark Matter Momenta}

The Lyman-$\alpha$ bounds of Table \ref{tab:masses} are model independent lower bounds on dark matter with a thermal momentum distribution, dependent only on   $g_{\rm DM}$. The reason for this is that one needs to match the relic density, this is determined by eq.~(\ref{eq:tdm}) via matching to late time cosmological observables. 

Although Lyman-$\alpha$ bounds can be relaxed if $g_{\rm DM}$ is large, cf.~(\ref{eq:eqgdm}), increasing the internal number of degrees of freedom of the dark matter simultaneously increases the available occupation levels in the Fermi gas and therefore shrinks the Fermi gas core size (cf.~eq.~(\ref{eq:scalingr0})). Changing the number of flavors of dark matter weakens the bound by $N_f^{-1/4}$, but the core radius also scales as $m N_f^{-1/4}$. Notably, these effects scale identically. Thus for larger $g_{\rm DM}$ one needs increasingly lighter dark matter to maintain appreciable cores, and larger cores ($R_c>0.2$~kpc) remain in tension with Lyman-$\alpha$. 

This conclusion can be circumvented if the dark matter momentum distribution is not thermal, but  skewed to lower energies. In this case the bounds from Lyman-$\alpha$ can be relaxed, potentially permitting lighter dark matter, and hence larger cores. Skewed momenta might arise in models with resonantly produced sterile neutrinos \citep{Shi:1998km} or preheating \citep{Kofman:1997yn}). 

In an appendix we also consider an alternative scenario in which the dark matter  has fewer effective degrees of freedom later in the cosmological evolution than when it was first produced. The model has $N_f$ flavors but also has small mass splittings that lead the heavier dark matter particles to decay without heating the remaining population. As a result of the large $N_f$, the temperature of the hidden sector can be lower while maintaining the relic DM density and at the same time appear to have fewer degrees of freedom during core formation and therefore lead to larger cores. If we choose $N_f = 16$, we can relax the Lyman-$\alpha$ mass bound by a factor of two down to $m>235$~eV and completely remove the tension between core sizes and Lyman-$\alpha$ mass constraints. For more details see Appendix~\ref{sec:skewed}.

\section{The Chandrasekhar limit}
\label{sec:chandra}

As one final consideration we look at core collapse into a black hole, and find that it does not
further constrain our system. Solving the Lane-Emden equation in the ultra-relativistic limit one obtains the Chandrasekhar bound for a given dark matter mass, as derived in \cite{Domcke:2014kla}
\beq
M_{\rm Ch}=8\pi^2 \sqrt{6} \omega^0_3 \frac{M_{\rm Pl}^3}{m^2} =  5\times10^{18}\left(\frac{\rm eV}{m}\right)^2M_\odot~,
\eeq
where $\omega^0_3\approx2.018$ is a numerical constant. 

Without the thermal envelope the solutions are similar to white dwarfs or neutron stars. These objects expel most of their outer layers during their formation and look like the solutions of \cite{Domcke:2014kla}.
The condition for the core of a dwarf galaxy of degenerate Fermi gas to collapse into a black hole is that the degenerate core is more massive than the Chandrasekhar limit for a given fermion mass. The mass enclosed in the core is
\beq
M_c\sim\rho_0 R_a^3\sim\frac{M_{\rm Pl}^3\sqrt{\rho_0}}{m^{4}}\left(\frac{3}{\pi N_f}\right)~,
\eeq
where a typical dwarf galaxy central density is of order $\rho_0 \sim 10^{-20} \text{kg/m}^3$. A dwarf galaxy core collapses into a black hole if $M_c > M_{\rm Ch}$, i.e. when
\beq
m \lesssim \rho_0^{1/4} \approx 0.1\;\text{eV}~.
\eeq
However, for $m \lesssim 0.1$ eV, the mass of the core, $10^{20} M_\odot$, exceeds the mass of any galaxy and therefore dwarf galaxies with degenerate fermion gas cores are never in danger of collapsing into black holes without additional mechanisms.



\section{Conclusion}
\label{sec:conc}
A (quasi) degenerate Fermi gas in dwarf galaxies can alter small scale structure formation. This can provide an elegant and minimal solution to the core-cusp problem.  Using the criterion that cores of order a  few 100 pc are required to reproduce the observations of dwarf galaxies, we have shown that suitable dwarf galaxy cores can be achieved for fermion dark matter with mass in the range 70~eV -- 400~eV. Ultralight fermions arise in many motivated contexts, such as gravitinos, or sterile neutrinos.

Modeling the dark matter as a quasi degenerate Fermi gas provides a well defined profile, eq.~(\ref{eq:scalingr0}) for describing the density distribution of dwarf galaxies, as illustrated in Figure \ref{fig:pclpf}. Here we have proposed a two-scale profile for general modeling of dwarf galaxy density profiles. We have argued that the two-scale profile defined in eq.~(\ref{eq:twosc}) may provide more physically motivated fits. This is because with only a single variable scale, as in Burkert and NFW profiles, fits to the dwarf galaxy profile are sensitive to the physics responsible for the core and the asymptotic behavior at large radial distances. Since the physics of the asymptotic behavior and the coring are likely unrelated it is important to treat these two length scales as independent.

Employing the Burkert profile, it was shown in \cite{Burkert:2015vla} that both large and dwarf galaxies obey a scaling relationship $\langle\rho_0 R_c\rangle= 75^{+85}_{-45}~M_{\odot}~{\rm pc}^{-2}$, see also \cite{Donato:2009ab}. The assumption of a quasi-degenerate Fermi gas profile automatically implies a relationship between the characteristic core size $R_c \sim R_0$. From eq.~(\ref{eq:scalingr0})  it follows that $\rho_0 R_c^{6}m_{\DM}^8 \sim {\rm constant}$. Interestingly, for dark matter in the mass range to give realistic core sizes, the scaling relationship identified by \cite{Burkert:2015vla} can be accommodated. This scaling relationship is illustrated in Figure \ref{fig:rhocvsrc}.

While sub-keV dark matter is typically excluded for a thermal relic, it is a viable possibility for non-thermal dark matter models \citep{Feng:2008mu,Berezhiani:1995am}, or Flooded Dark Matter \citep{Randall:2015xza}. Dark matter can become non-relativistic earlier in these models which would weaken the limits from small scale structure observations compared to thermal relics \citep{Bolton:2004ge,Viel:2013fqw}. Moreover, we argued that if the dark matter momentum distribution is non-thermal and skewed to lower energies, then the Lyman-$\alpha$ bounds are relaxed. In the case that free streaming bounds can be ignored, the leading conservative lower bound is $m\gtrsim 70$ eV, derived here in Section~\ref{sec:lowermassbounds}.  

We have focused on the case that Fermi repulsion is entirely responsible for coring dwarf galaxies, and resolving the core-cusp problem. However in principle the cores in dwarfs could emerge due to a combination of Fermi repulsion and baryonic feedback. In the case that baryonic feedback plays a role, this might allow for appreciable cores for somewhat heavier dark matter.  It would be interesting to combine these effects in a numerical simulation.

\vspace{-4mm}\subsection*{Acknowledgements}

We are grateful to P.~Agrawal, N.~Amorisco, J.~Binney, F.~Y.~Cyr-Racine, M.~Reece, A.~Urbano, M.~Walker, E.~Witten, and I.~Yavin for useful discussions.  This work was supported in part by NSF grants PHY-0855591, PHY-1216270, and PHY-1415548, and the Fundamental Laws Initiative of the Harvard Center for the Fundamental Laws of Nature. 

\appendix

\section{The Plummer Profile}
\label{ApB}

For completeness we give here the profiles and functions used in the
definition of $\sigma_{\sc LOS}$
 in eq.~(\ref{slos}). The Plummer profile \citep{Plummer:1911zza} for
the projected stellar density $I(r)$ is defined, in terms of the
total luminosity $L$, as follows
\beq
I(r)=\left[1+\left(\frac{r}{r_{\nicefrac{1}{2}}}\right)^2\right]^{-2}\left(\frac{L}{\pi
r_{\nicefrac{1}{2}}^2}\right)~.
\label{Ir}
\eeq
Given eq.~(\ref{Ir}) the associated 3D  density is  \citep{Binney}
\beq
\nu(r')=\left[1+\left(\frac{r'}{r_{\nicefrac{1}{2}}}\right)^2\right]^{-\nicefrac{5}{2}}\left(\frac{3L}{4\pi
r_{\nicefrac{1}{2}}^3}\right)~.
\eeq
The $\beta$-dependent function which appears in eq.~(\ref{slos}) is
\beq
F(\beta,r,r')=\int_r^{r'}\left[1-\beta\left(\frac{r'}{r''}\right)^2\right]\frac{(r'')^{1-2\beta}}{\sqrt{r''{}^2-r^2}}~{\rm
d}r''.
\eeq

\section{Models of Dark Matter}
\label{Apmodels}
\subsection{Flooded Dark Matter}
\label{sec:FDM}

Cosmological bounds on light dark matter can be relaxed if the dark matter sector is colder than the Standard Model sector. This scenario arises in Flooded Dark Matter  \citep{Randall:2015xza} but includes other scenarios as well. We first derive a bound on the temperature ratio that allows for thermal populations to give the correct energy and number densities that applies to any model.
We allow for decoupling but assume that  dark matter is nonetheless of thermal origin and its density is $n \propto T_{\rm DM}^3$, where $T_{\rm DM}$ is its temperature.  The density of dark matter today is then
\beq
\Omega_\DM = m_\DM n_\DM = m_\DM g_\DM T_{\DM,0}^3~,
\eeq
where $T_{\DM,0}$ denotes the temperature today.
Similarly, for the Standard Model sector
\beq
\Omega_{\rm B} = m_N \eta g_\gamma T_{\SM,0}^3~.
\eeq
Here $m_N$ is the mass of a nucleon and $\eta = n_B/s = 6.2\times 10^{-10}$ is the baryon asymmetry. Assuming dark matter froze out while relativistic, the ratios of temperatures of the dark and Standard Model sectors must be related by:
\beq
\frac{T_{\DM}}{T_{\SM}} = \left( \frac{g_{\gamma}}{g_{\rm DM}}  \eta
\frac{\Omega_{\rm DM}}{\Omega_{B}} \frac{m_N}{m_{\rm DM}}\right)^{1/3}.
\label{eq:tdmtsm}
\eeq
For dark matter masses above an eV, the dark matter sector can be colder than the Standard Model, in which case the Lyman-$\alpha$ bound on the dark matter mass would be weaker. This relation applies to any model with a thermal population of dark matter. To avoid this conclusion requires a nonthermal distribution.

This is also the conclusion in the Flooded Dark Matter scenario \citep{Randall:2015xza}.
Suppose the dark matter constitutes a hidden sector which is essentially decoupled from the visible sector, and that after inflation these sectors are reheated democratically. To reproduce standard cosmology at late time the visible sector must typically receive further entropy injections. A long lived heavy state $\Phi$ which decays only to the visible sector can provide an appropriate source of entropy production. The evolution of the light states and $\Phi$ are distinct. $\Phi$ becomes non-relativistic at early time, at which point its energy density $\rho_\Phi$ redshifts as matter. Provided $\Phi$ is sufficiently long-lived, it will come to dominate the energy density of the Universe. 

When $\Phi$  decays its energy density is transferred to the Standard Model states. The decay rate required to match the observed dark matter relic density ($\Omega_{\rm DM}\sim5\Omega_B$) is
\beq
 \Gamma \sim \frac{m_\Phi^2}{M_{\rm Pl}}\left(\frac{g_{\gamma}}{g_{\rm DM}}  \eta \frac{ \Omega_\DM}{\Omega_B}\frac{m_N}{m_{\rm DM}}\right)^2~.
\label{eq:Ga-0}
\eeq
The entropy injection from $\Phi$ decays heat up the visible sector and sets the temperature ratio as required by eq.~\ref{eq:tdmtsm}.
Diluting the dark matter sufficiently requires small $\Phi$ decay rates, as dictated by eq.~(\ref{eq:Ga-0}). However successful Big Bang Nucleosynthesis sets a lower bound on the reheat temperature $T_{\rm RH}\simeq\sqrt{\Gamma M_{\rm Pl}}\gtrsim10~{\rm MeV}$. Moreover, certain baryogenesis scenarios need higher $T_{\rm RH}$, in particular mechanisms tied to the electroweak phase transition require $T_{\rm RH}\gtrsim100$ GeV.  See e.g.~\cite{Morrissey:2012db} for models and constraints.

\subsection{Skewed Dark Matter Momenta}
\label{sec:skewed}

Consider $N_f$ dark matter `flavors' of 200-300 eV dark matter with a thermal momentum distribution at $T_{\rm DM}<T_{\rm SM}$. To avoid the Lyman-$\alpha$ limit one needs $N_f\sim16$ (cf.~Table \ref{tab:masses}). Even lighter dark matter can be accommodated with larger $N_f$. 
Moderate values of $N_f$ could arise due to an adjoint representation of some broken SU($\sqrt{N_f}$), and could be connected to ideas of `Flavored Dark Matter' \citep{Agrawal:2011ze,Batell:2011tc}. We envisage that the a weakly gauged dark sector gauge group breaks, yielding a low energy theory consisting of $N_f$ closely spaced fermions with small mass splittings $\delta_{ij}$. If the gauge symmetry is broken by high scale operators, cutoff at $\Lambda$, then the induced mass splittings are parametrically 
\beq
\delta_{ij}\equiv\frac{m_{\chi_i}-m_{\chi_j}}{m_{\chi_0}}\sim\frac{\lambda_X^2}{16\pi^2}{\rm Log}\left[\frac{m_{\chi_0}}{\Lambda}\right]~,
\eeq  
where $\chi_0$ is the lightest state and $\lambda_X$ is the gauge coupling. 

Eventually the (slightly) heavier states $\chi_i$ decay to light hidden sector states $X$, removing the $\chi_i$ population. In order to relax the Lyman-$\alpha$ bounds, $\chi_i$ decays should occur after the dark matter decouples from the lighter species. Otherwise the $\chi_0$ would receive a compensating energy contribution from the $\chi_i$ decays through interactions with the $X$-bath. 
As a result there is a reduction of the number of degrees of freedom,  but due to the small mass splitting, there is no appreciable heating of the $\chi_0$ population. Accordingly the momentum distribution of the $\chi_0$ states appears with a temperature which is scaled relative to the thermal expectation by a factor of $g^{1/4}=(2N_f)^{1/4}$. 

The $\chi_0$ distribution is modified as follows:
\beq
df(p) = dp^3\left[1+ \exp \left( \frac{(2N_f)^{1/4} p }{ T} \right)\right]^{-1},
\eeq
 where $T$ here can be identified with $T_{\rm DM}$ in eq.~(\ref{eq:tdmtsm}). 

As indicated above, in order for the heavier $\chi_i$ states to decay the model must include light or massless hidden sector states $X$. The energy injected into these radiation states will subsequently redshift away. Since the hidden sector is cold relative to the visible sector the $X$ states will typically not give an appreciable contribution to the number of relativistic species $N_{\rm eff}$ \citep{Randall:2015xza}. Generally, this decay should occur with a small coupling to avoid
a) thermalization and hence reheating of the remaining light states, b) freeze-out, and c) self-scattering effects which are not ruled out but would greatly alter our core-cusp analysis.

Specifically, for  too large $\lambda_X$, dark matter annihilations can be significant and there can be a stage of freeze-out dynamics which would impact the dark matter relic density. 
 Therefore to avoid $\chi_0$ freeze-out the annihilations $\chi_0\chi_0\rightarrow XX$ must be inactive down to the decoupling temperature: $T\sim m_{\chi_0}$. Given that
the annihilation cross section of $\chi_0$ is parametrically  $\langle \sigma v\rangle\sim \lambda_X^4/16\pi m_{\chi_0}^2$, which should be insignificant down to the decoupling temperature, we require that
\beq
\lambda_X \lesssim \left(\frac{8\pi m_{\chi_0}}{M_{\rm Pl}}\right)^{1/4} \sim 10^{-6}
\eeq
Furthermore, to avoid significant self scattering we require \citep{Peter:2012jh}
\beq
\lambda_X^4/(v^4 m_{\chi_0}^3) \lesssim 0.1 {\rm cm}^2/{\rm g}
\eeq
For velocity dispersions of $\sim$10 km/s, this gives $\lambda_X \lesssim 10^{-9}$.

\label{lastpage}

\end{document}